# Nuclear Shell Structure Evolution Theory

Zhengda Wang[1], Xiaobin Wang[2], Xiaodong Zhang[3], Xiaochun Wang[3]

(1) Institute of Modern physics, Chinese Academy of Sciences，Lan Zhou,

P. R. China 730000

(2) Seagate Technology, 7801 Computer Avenue South, Bloomington, Minnesota

U.S.A. 55435

(3) University of Texas, M. D. Anderson Cancer Center, Radiation Physics

Department, Box 1202, Houston, Texas, U.S.A. 77030

First draft in Nov, 2010

# Abstract

The Self-similar-structure shell model (SSM) comes from the evolution of the conventional shell model (SM) and keeps the energy level of SM single particle harmonic oscillation motion. In SM, single particle motion is the positive harmonic oscillation and in SSM, the single particle motion is the negative harmonic oscillation. In this paper a nuclear evolution equation (NEE) is proposed. NEE describes the nuclear evolution process from gas state to liquid state and reveals the relations among SM, SSM and liquid drop model (DM). Based upon SSM and NEE theory, we propose the solution to long-standing problem of nuclear shell model single particle spin-orbit interaction energy $\varepsilon_{nlj}^{s \cdot L}$. We demonstrate that the single particle motion in normal nuclear ground state is the negative harmonic oscillation of SSM[1][2][3][4]

# I. Introduction

For nucleus at ground state, because nucleon density saturates at the center region, there is a potential formed at the center region for the saturated nuclear force. Nuclear shell model single particle mean field potential is not the positive harmonic oscillator potential in SM, but the negative harmonic oscillator potential as in SSM. For $N^{15}$ and $O^{15}$, the experimental energy level of the single particle harmonic oscillation between $\frac{1}{2}-$ and $\frac{3}{2}+$ is about: $\hbar\omega^{ex}(N^{15},O^{15}) \approx 7.0 MeV$ [5]. Based upon conventional SM approach, this energy level could be derived from nuclear liquid drop radii $R_{DM}$: $\hbar\omega^{SM} \approx 41A^{-1/3} MeV$, where $R_{DM} = r_0 A^{1/3}, r_0 = 1.2 fm$. However, for ($N^{15}, O^{15}$), the derived SM energy level interval is $\hbar\omega^{sm}(N^{15},O^{15}) \approx 16 MeV$, much larger than the experiment value. According to SM physics picture, the smaller $\hbar\omega^{ex}(N^{15},O^{15})$ nuclei should be formed at a state with radii R($N^{15},O^{15}$) much larger than the liquid drop radii $R_{DM}(N^{15},O^{15})$: R($N^{15},O^{15}$) >> $R_{DM}$ ($N^{15},O^{15}$).

The gas state nucleus radius $R_{gas}$ is much larger than liquid drop radius $R_{DM}$. For gas state nucleus, nuclear density is not saturated at the center region. The nuclear force is not saturated and the single particle mean field potential is well described by the SM positive harmonic oscillator potential. The experiment observed smaller $\hbar\omega^{ex}(N^{15},O^{15})$ nuclei are formed under the gas state condition, according to SM picture. $(N^{15}, O^{15})$ nuclei originally come from nuclear gas generated by big-bang. $\hbar\omega^{ex}(N^{15}, O^{15})$ generated at the gas state is not changed during nuclear evolution liquefaction process. This will be further discussed in later sections.

During nuclear evolution from gas state to liquid state, nucleon density at the center region increases. The initial unsaturated nucleon density evolves toward saturation. The nuclear shell model single particle mean field potential should be changed from SM positive harmonic oscillator potential to SSM negative harmonic oscillator potential to describe nuclear evolution process. Although traditional SM does not consider nuclear evolution process, it provides the best starting point in the study of nuclear shell structure evolution. In the study of nuclear evolution, the nuclear state corresponding to "border nucleus" is very important. At "border nucleus" state, the shell structure single particle motion changes from the SM positive harmonic oscillation to the SSM negative harmonic oscillation. The nuclear evolution process before "border nucleus" is described by SM gas state and the nuclear evolution process after "border nucleus" is described by SSM liquid state. Border nucleus is the real halo nucleus. In the following, we will obtain our conclusions through study of SM, DM, SSM and NEE and verify the theoretical prediction on existing experiment measurements.

## II. Self Similar Structure Shell Model (SSM) and Spin-Orbit Interaction Energy

According to the conventional point of view, adding a negative constant $V_C$ to the single particle harmonic oscillator potential $U_{os}(r)$ does not change the mathematical solution of single particle Hamiltonian $H_0^{sm}$ in SM. However, in SSM, through rescaling harmonic oscillator Hamiltonian with a negative potential energy $V_c$, the single particle positive harmonic oscillator Hamiltonian is transformed to the negative harmonic oscillator Hamiltonian $H_0^{SSM}$. This transformation keeps single particle motion energy levels and configuration combinations unchanged. When nucleon density at the center is not saturated, SM positive harmonic oscillation motion is the correct solution for nuclear shell structure. When nucleon density saturates at the center region, SSM negative harmonic oscillation motion is the correct solution for nuclear shell structure.

In SM the positive harmonic oscillation of single particle is described by the Hamiltonian $H_0^{SM}$ [1].

$$H_0^{SM} = H_0 + V_c \tag{1}$$

$$H_0 = -\frac{\hbar^2}{2m}\left(\frac{d^2}{dx^2}+\frac{d^2}{dy^2}+\frac{d^2}{dz^2}\right)+\frac{m\omega^2}{2}\left(x^2+y^2+z^2\right) \tag{2}$$

$$H_0^{SM}\psi_{nl}^{SM}(\alpha r,\theta,\varphi)=\left[(n+\frac{3}{2})\hbar\omega+V_c\right]\psi_{nl}^{SM}(\alpha r,\theta,\varphi)$$

$$=-\left[\frac{-V_c}{(n+\frac{3}{2})\hbar\omega}-1\right]H_0\psi_{nl}^{SM}(\alpha r,\theta,\varphi) \quad (3)$$

$$=-c_{nl}^2\,\psi_{nl}^{SM}(\alpha r,\theta,\varphi)$$

$$c_{nl}^2=\sqrt{\left[\frac{-V_c}{(n+\frac{3}{2})\hbar\omega}-1\right]} \quad (4)$$

$$H_0^{SM}\psi_{nl}^{SM}(\alpha r,\theta,\varphi)=\varepsilon_{nl}^{SM}(0)\psi_{nl}^{SM}(\alpha r,\theta,\varphi) \quad (5)$$

$$\varepsilon_{nl}^{SM}(0)=\left(n+\frac{3}{2}\right)\hbar\omega+V_c \quad (6)$$

$$\alpha=\sqrt{\frac{m\omega}{\hbar}} \quad (7)$$

In SSM the negative harmonic oscillation of single particle is described by the Hamiltonian $H_0^{SSM}$ [2].

$$\begin{aligned}H_0^{SSM}&=-c_{nl}^2 H_0\\ &=-c_{nl}^2\left[-\frac{\hbar^2}{2m}\left(\frac{d^2}{dx^2}+\frac{d^2}{dy^2}+\frac{d^2}{dz^2}\right)+\frac{m\omega^2}{2}\left(x^2+y^2+z^2\right)\right]\\ &=-\left[-\frac{\hbar^2}{2m}\left(\frac{d^2}{d\left(\frac{x}{c_{nl}}\right)^2}+\frac{d^2}{d\left(\frac{y}{c_{nl}}\right)^2}+\frac{d^2}{d\left(\frac{z}{c_{nl}}\right)^2}\right)+\frac{m(c_{nl}^2\omega)^2}{2}\left(\left(\frac{x}{c_{nl}}\right)^2+\left(\frac{y}{c_{nl}}\right)^2+\left(\frac{z}{c_{nl}}\right)^2\right)\right]\end{aligned} \quad (8)$$

$$\omega_{nl}^{SSM}=c_{nl}^2\omega \quad (9)$$

$$\alpha_{nl}=c_{nl}\alpha \quad (10)$$

$$H_0^{SSM}\psi_{nl}^{SSM}(\alpha_{nl}r,\theta,\varphi)=\varepsilon_{nl}^{SSM}(0)\psi_{nl}^{SSM}(\alpha_{nl}r,\theta,\varphi) \quad (11)$$

$$\varepsilon_{nl}^{SSM}(0)=\varepsilon_{nl}^{SM}(0)=\left(n+\frac{3}{2}\right)\hbar\omega+V_c \quad (12)$$

Conventional SM neglects $V_c$, thus could not obtain the negative harmonic oscillation motion as in SSM. The energy level differences in SSM are the same as those of the SM, However, $\hbar\omega_{nl}^{SSM}$ in SSM is different from $\hbar\omega$ in SM. This needs to be pointed out specifically. In SSM, the single particle negative oscillation circular frequency $\omega_{nl}^{SSM}$ depends upon the single particle configuration. It is not a constant as in SM. This makes nucleon spatial probability distribution more reasonable. The physics picture of SSM single particle negative harmonic oscillation motion in image space (ir , ip) is similar to the physical picture of SM positive harmonic oscillation motion in real space (r, p). While the real space corresponds to positive energy space the image space corresponds to negative energy space.

SSM spin-orbit interaction energy $\varepsilon_{nlj}^{s\cdot L}(SSM)$ is:

$$\begin{aligned}\varepsilon_{nlj}^{s\cdot L}(SSM) &= 3^2\left[\frac{1}{2mc^2}\frac{1}{r}\frac{dU_{os}(nl,r)}{dr}\right]\vec{s}\cdot\vec{L} \\ &= -\frac{3^2(\omega_{nl}^{SSM}\hbar)^2 n_{ls}}{2mc^2\hbar^2}\end{aligned} \quad (13)$$

$$n_{ls} = \frac{l}{2}\hbar^2 \quad \text{for } j = l+1/2 \quad (14)$$

$$n_{ls} = -\frac{(l+1)}{2}\hbar^2 \quad \text{for } j = l-1/2 \quad (15)$$

where $U_{os}^{SSM}(nl,r)$ is the single particle negative harmonic oscillator potential:

$$U_{os}^{SSM} = -\frac{m(\omega_{nl}^{SSM})^2 r^2}{2} \quad (16)$$

Above formula $\varepsilon_{nlj}^{s\cdot L}(SSM)$ without $3^2$ factor is derived directly from the relativity quantum mechanics. This spin-orbit interaction formula has been successfully used in describing atom shell structure. $3^2$ factor comes from the consideration of a single nuclear particle consisting of three quarks. Three quarks take part in the spin-orbit interaction. One single particle orbit corresponds to three quark orbits and one single particle spin corresponds to three quarks spins.

Although there is no reason to believe that quantum mechanics spin-orbit interaction formula $\varepsilon_{nlj}^{s \cdot L}$ could not be used in nuclear shell model, the spin-orbit interaction of single particle motion in nuclear shell model has not been solved for long time. There are two reasons that $\varepsilon_{nlj}^{s \cdot L}$ formula in SM is not successful. Firstly because SM single particle mean field is positive harmonic oscillator potential, $\varepsilon_{nlj}^{s \cdot L}$ with $j = l + 1/2$ has positive value and $\varepsilon_{nlj}^{s \cdot L}$ with $j = l - 1/2$ has negative value. The sign obtained from $\varepsilon_{nlj}^{s \cdot L}(SM)$ is opposite to the experiment value $\varepsilon_{nlj}^{s \cdot L}(EXP)$ for normal nuclei. The sign of $\varepsilon_{nlj}^{s \cdot L}(SM)$ is the same to the sign of electron orbit coupling energy in atom shell structure. Secondly, the conventional SM $\varepsilon_{nlj}^{s \cdot L}(SM)$ formula does not consider that a single particle (nucleon) consists of three quarks. $3^2$ factor is not included in the conventional SM spin-orbit interaction formula. Thus the calculated SM $\varepsilon_{nlj}^{s \cdot L}(SM)$ value is much smaller than that of the experiment value $\varepsilon_{nlj}^{s \cdot L}(EXP)$. In SSM, the single particle mean field is negative harmonic oscillator potential. The sign between $\varepsilon_{nlj}^{s \cdot L}(SSM)$ and $\varepsilon_{nlj}^{s \cdot L}(SM)$ is exactly opposite. $\varepsilon_{nlj}^{s \cdot L}(SSM)$ has the same sign as experiment measurement $\varepsilon_{nlj}^{s \cdot L}(EXP)$ for normal nucleus. Also here we consider that a single nuclear particle consists of three quarks and three quarks take part in the spin-orbit interaction directly. In nuclear shell structure, the two single particle states of $j = l \pm 1/2$ corresponds to the two quark states of $j = l \pm 1/2$ and the quark state has $3^2$ degeneracy. Also $\omega_{nl}^{SSM}$ in SSM depends on single particle configuration, the problem of $\varepsilon_{nlj}^{s \cdot L}$ changing with orbit angular momentum is solved naturally. In the later section of calculation and discussions, it will be shown that calculated $\varepsilon_{nlj}^{s \cdot L}(SSM)$ fits very well to experiment measurements $\varepsilon_{nlj}^{s \cdot L}(EXP)$. This validates the negative harmonic oscillation motion in SSM.

SSM spin-orbit interaction energy $\varepsilon_{nlj}^{s \cdot L}(SSM)$ can be written as:

$$\varepsilon_{nlj}^{s \cdot L}(SSM) = c_{nl}^{s \cdot L}(SSM)\vec{s} \cdot \vec{L} \tag{17}$$

$$c_{nl}^{s \cdot L}(SSM) = \frac{-3^2}{2mc^2\hbar}\left(\omega_{nl}^{SSM}\hbar\right)^2 \tag{18}$$

SSM energy correction term $\varepsilon_{nl}^{L}(SSM)$ due to $\hat{L}^2$ can be written in a

conventional format:

$$\varepsilon_{nl}^{L}(SSM) = D_{nl}^{L}(SSM)\hat{L}^2 \tag{19}$$

$$D_{nl}^{L}(SSM,n) = -\frac{1}{(1+e^{14-6n})}\frac{\hbar\omega_{bn}^{SM}}{(n+\frac{3}{2})^2} \tag{20}$$

$$D_{nl}^{L}(SSM,p) = -\frac{1}{(1+e^{19.45-6n})}\frac{\hbar\omega_{bp}^{SM}}{(n+\frac{3}{2})^2} \tag{21}$$

where $\hbar\omega_{b}^{SM}$ corresponds to SM border nucleus single particle positive harmonic oscillation energy level. This can be calculated theoretically by nuclear evolution equation (NEE), which will be discussed in details in the next section.

In summary, SSM single particle motion Hamiltonian $H^{SSM}$ is:

$$H^{SSM} = H_0^{SSM} + \varepsilon_{nlj}^{s\cdot L}(SSM) + \varepsilon_{nl}^{l}(SSM) \tag{22}$$

$$H^{SSM}\psi_{nlj}^{SSM}(\alpha_{nl}r,\theta,\varphi) = \varepsilon_{nlj}^{s\cdot L}(SSM)\psi_{nlj}^{SSM}(\alpha_{nl}r,\theta,\varphi) \tag{23}$$

$$\varepsilon_{nlj}^{SSM} = \varepsilon_{nl}^{SSM}(0) + \varepsilon_{nlj}^{s\cdot L}(SSM) + \varepsilon_{nl}^{l}(SSM) \tag{24}$$

### III. Nuclear Evolution Equation (NEE)

Nuclear evolution equation NEE describes nuclear evolution from gas state to liquid state. It reveals the relation among SM, SSM and DM. Because the single particle motion velocity is much larger than the nuclear evolution velocity, the probability distribution of single particle motion in nuclear structure can be treated quasi-statically. The average of kinetic energy is equal to the average of potential energy for single particle harmonic oscillation and is half of the single particle energy. The summation of interacting energies among all nucleons in nucleus is equal to a half of nuclear binding energy. Nuclear binding energy is given by DM model $E_{DM}$. Following relation holds between SM single particle energy $\varepsilon_{nlj}^{SM}$ and DM energy $E_{DM}(A,N)$:

$$\frac{1}{2}\sum_{1}^{A}\left[\varepsilon_{nl}^{SM}(0)+2\varepsilon_{nlj}^{s\cdot L}+\varepsilon_{nl}^{l}\right]=-E_{DM}(A,N) \quad (25)$$

$$E_{DM}(A,N)=a_{v}A-a_{s}A^{2/3}-a_{c}Z^{2}A^{-1/3}-a_{\alpha}\left(\frac{A}{2}-Z\right)^{2}A^{-1}+a_{P}\delta A^{-1/2} \quad (26)$$

$a_{v}=15.835 MeV, a_{s}=18.33 MeV$, $a_{c}=0.714 MeV, a_{\alpha}=92.8 MeV, a_{p}=11.2 MeV$ and $\delta=1$ for even even nuclei, $\delta=0$ for odd nuclei and $\delta=-1$ for odd odd nuclei. Equation (25) is the SM nuclear evolution equation NEE(SM). The $1/2$ factor in NEE (SM) is counting the repeated summation of interactions among all nucleons.

In $\varepsilon_{nlj}^{SM}$, single particle energy $\varepsilon_{nlj}^{SM}(0)$ has two parts. The first part is the potential well depth $V_c$ and the other part is the energy associated with positive harmonic oscillating motion $\hbar\omega^{SM}$ (eq(6)). As NEE shows, due to the conserved energy, A decreasing in $V_c$ results an increasing in $\hbar\omega^{SM}$.

This reduces the mean square root radius $r_{nl}$ of single particle motion and increases nucleon density:

$$r_{nl}^2 = \frac{\left(n + \frac{3}{2}\right)\hbar}{m\omega^{SM}} \quad (27)$$

NEE(SM) describes the nuclear shell structure evolution process in gas state in terms of $V_c$ and $\hbar\omega^{SM}$. Conventional SM neglects $V_c$ term and could not describe nuclear evolution process. SM and DM describe nuclear ground state, the lowest energy level of an isolated system. Nuclear evolution process described by NEE(SM) is a result of energy conservation.

When $V_c$ decreases to the double volume energy $-2a_v$ in DM, $Vc = -2a_v$, the nucleus reaches "border nucleus" state. The single particle motion in shell model changes suddenly from the SM positive harmonic oscillation to the SSM the negative harmonic oscillation. Border nucleus is at the end of SM evolution process and at the start of SSM evolution process. In nuclear evolution process from gas state to liquid state, the shell single particle mean field potential transforms from the SM positive harmonic oscillator potential to the SSM negative harmonic oscillator potential. The reasons for nuclear single particle motion to change suddenly from the positive harmonic oscillation to the negative harmonic oscillation at "border nucleus" $Vc = -2a_v$ are as follows: (1) Because the nucleon density and nuclear force at nuclear center region start to saturate at border nucleus, a negative harmonic oscillator potential in SSM starts to form. (2) NEE(SM) energy relation between SM and DM for border nucleus can be found as:

$$-a_v A = \frac{1}{2} V_c A \tag{28}$$

$$a_s A^{2/3} + a_c Z^2 A^{-1/3} + a_\alpha \left(\frac{A}{2} - Z\right)^2 A^{-1} - a_P \delta A^{-1/2}$$
$$= \frac{1}{2} \sum \left[ (n + \frac{3}{2}) \hbar \omega_b^{SM} + 2\varepsilon_{nlj}^{s \cdot L} + \varepsilon_{nl}^l \right] \tag{29}$$

The depth $V_c$ of potential well in SM corresponds to the negative volume energy $(-a_v)$ of DM. The summation of all single particle harmonic oscillation energies in SM corresponds to the positive energy summation in DM (i.e. Coulomb energy plus surface energy plus symmetric energy). If we insisted to describe nuclear evolution process after border nucleus as gas state in SM, the energy relation in eq (28) and eq (29) between SM and DM would be destroyed and the contradiction between SM and DM would be resulted. (3) As mentioned in the introduction, if nuclear evolution process after border nucleus were SM gas state, the $\hbar\omega^{SM}$ would increase and the calculated value of $\hbar\omega^{SM}$ would not be consistent with experiment value $\hbar\omega^{EXP}(N^{15},O^{15}) \ll \hbar\omega^{SM}(N^{15},O^{15})$. (4) As discussed in the previous section of spin orbit interaction energy $\varepsilon_{nlj}^{s \cdot L}$, if nuclear evolution process after border nucleus were still SM gas state, the theoretical values of $\varepsilon_{nlj}^{s \cdot L}(SM)$ would not be consistent with experiment measurements $\varepsilon_{nlj}^{s \cdot L}(EXP)$. (5) At border nucleus, all the energies are converted to the SSM negative harmonic oscillating energy, the absolute value of summation of all single particle positive harmonic oscillation energies is about equal to the absolute value of summation of all negative harmonic oscillation energies, $\sum(n+\frac{3}{2})\hbar\omega_{nl}^{SSM} \approx \sum(n+\frac{3}{2})\hbar\omega_b^{SM}$ and $c_{nl}^2 \approx 1$. These five reasons are also the reasons why we believe single particle motion in normal nucleus at ground state is the negative harmonic oscillation as in SSM. In conventional SM, $V_C$ is set to be zero. There is no nuclear evolution equation NEE(SM) and the energy relation between SM and DM (eq (28) and eq (29)) can not be obtained.

In SSM liquefaction process, the following conclusions can be obtained: (1) As $V_c$ decreases, the single particle $\hbar\omega^{SSM}$ in SSM increases (eq(4) and eq(9)) and the single particle radius decreases (eq(27)). (2) In SSM liquefaction process, further decreasing $V_c$ will not change the orders and intervals between single particle energy levels $\varepsilon_{nl}^{SSM}(0)$ left at nucleus border. (3) In SSM liquefaction process, all the negative energy associated with $V_c$ decreasing are converted to negative harmonic oscillating energies. The conservation of energy for nucleus system requires nuclear temperature increasing to cancel the negative energy $V_c$ decreasing:

$$T = -(V_c + 2a_v) \tag{30}$$

where the border nucleus temperature is defined to be zero $T_b = 0$. The temperature increasing energy comes from the liberation energy in liquefaction process.

The decreasing of $V_c$ in the liquefaction process results the increasing of $\hbar\omega^{SSM}$ and T. This continues until the nucleus is fully liquefied to form normal nucleus at ground state. The nuclear evolution equation NEE (SSM) corresponding to SSM liquefaction process is:

$$\frac{1}{2}\sum_{l=1,A}\left[\varepsilon_{nl}^{SSM}(0) + 2\varepsilon_{nlj}^{s\cdot L} + \varepsilon_{nl}^{l} + T\right] = -E_{DM}(A,N) \tag{31}$$

There is the relation between SM and SSM:

$\varepsilon_{nl}^{SSM}(0) + T = \varepsilon_{nl}^{SM}(0, nl)$, $\varepsilon_{nl}^{SM}(0, nl)$ is $\varepsilon_{nl}^{SM}(0)$ at "border nucleus". In the SSM liquefaction process, the energy relation between SSM and DM is:

$$-a_v A = \frac{1}{2}(V+T)_c A \tag{32}$$

$$a_s A^{2/3} + a_c Z^2 A^{-1/3} + a_\alpha \left(\frac{A}{2} - Z\right)^2 A^{-1} - a_P \delta A^{-1/2}$$
$$= \frac{1}{2} \sum \left[(n+\frac{3}{2})\hbar\omega_b^{SM} + 2\varepsilon_{nlj}^{s\cdot L}(SSM) + \varepsilon_{nl}^{l}(SSM)\right] \tag{33}$$

The energy relation between SSM and DM in the SSM process is similar to the energy relation between SM and DM at border nucleus. The energy relationships eq(32) and eq(33) between SSM and DM show that SSM and DM are unified. The SSM is a quantum liquid drop model (QDM). In SSM process, the nuclear temperature T reflects the disorder motion of nucleons in nucleus. This provides insight to understand the physics picture of DM.

### IV. Calculations and Discussions

We can test nuclear evolution equation NEE and $\varepsilon_{nlj}^{SSM}$ formula through analyzing experiment measurements of single particle levels. For $N^{15}$ and $O^{15}$, the experiment data of single particle oscillation levels are[5]:

$\varepsilon_{11\frac{1}{2}}^{ex}(N^{15}, -\frac{1}{2}) = 0 MeV$, $\varepsilon_{22\frac{5}{2}}^{ex}(N^{15}, \frac{5}{2}) = 5.2704 MeV$, $\varepsilon_{22\frac{3}{2}}^{ex}(N^{15}, \frac{3}{2}) = 7.3011 MeV$,

$\varepsilon^{ex}_{11\frac{1}{2}}(O^{15}, -\frac{1}{2}) = 0 MeV$, $\varepsilon^{ex}_{22\frac{5}{2}}(O^{15}, \frac{5}{2}) = 5.2409 MeV$, $\varepsilon^{ex}_{22\frac{3}{2}}(O^{15}, \frac{3}{2}) = 6.79 MeV$,

$\hbar\omega(N^{15}) = \varepsilon^{ex}_{22\frac{3}{2}}(N^{15}, \frac{3}{2}) - \varepsilon^{ex}_{22\frac{1}{2}}(N^{15}, -\frac{1}{2}) = 7.3 MeV$,

$\hbar\omega(O^{15}) = \varepsilon^{ex}_{22\frac{3}{2}}(O^{15}, \frac{3}{2}) - \varepsilon^{ex}_{22\frac{1}{2}}(O^{15}, -\frac{1}{2}) = 6.79 MeV$,

For $N^{15}$ and $O^{15}$, $\hbar\omega_b^{SM}$ can be calculated by NEE(SM) at border nucleus ($V_c$=-2$a_v$), $\hbar\omega_b^{SM}(N^{15}) = 7.3 MeV$ and $\hbar\omega_b^{SM}(O^{15}) = 7.5 MeV$. The calculated $\hbar\omega_b^{SM}(N^{15}, O^{15})$ is close to experiment measurement values $\hbar\omega^{ex}(N^{15}, O^{15})$: $\hbar\omega_b^{SM}(N^{15}, O^{15}) \approx \hbar\omega^{ex}(N^{15}, O^{15})$. This is an important validation to NEE(SM, SSM). As pointed out in the introduction, conventional SM predicts $\hbar\omega^{SM}(N^{15}, O^{15})$ about two times bigger than experiment measurements $\hbar\omega^{ex}(N^{15}, O^{15})$.

Single particle spin-orbit interaction $\varepsilon^{s \cdot L}_{nlj}$ is a direct test to determine whether single particle motion in normal nucleus is SM positive harmonic oscillation or SSM negative harmonic oscillation. For $N^{15}$ $\varepsilon^{s \cdot L}_{22\frac{3}{2}}(N^{15}, ex) = -0.81 MeV$, $\varepsilon^{s \cdot L}_{22\frac{5}{2}}(N^{15}, ex) = +1.22 MeV$ can be obtained from energy difference $\varepsilon^{ex}_{22\frac{3}{2}}(N^{15}, \frac{3}{2}) - \varepsilon^{ex}_{22\frac{5}{2}}(N^{15}, \frac{5}{2})$. According to $\varepsilon^{s \cdot L}_{nlj}(SSM)$ formula in this paper, $\hbar\omega^{SSM}_{22} = 13.02 MeV$ and root mean square radius $r_{22\frac{3}{2}} = r_{22\frac{5}{2}} = 3.31 fm$ can be derived. Root mean square radius $r_{22\frac{3}{2}}$ and $r_{22\frac{5}{2}}$ are little larger than $R_{DM}$=2.96fm. Because $\varepsilon^{ex}_{22\frac{3}{2}}(\frac{3}{2})$ and $\varepsilon^{ex}_{22\frac{5}{2}}(\frac{5}{2})$ are single particle excited states at d shell (which is one shell above p shell), it is reasonable that calculated root mean square radius is a little bit larger than liquid drop radius. For $O^{15}$, $\varepsilon^{s\cdot L}_{22\frac{5}{2}}(O^{15},ex)=-0.62MeV$, $\varepsilon^{s\cdot L}_{22\frac{3}{2}}(O^{15},ex)=+0.93MeV$, $\hbar\omega^{SSM}_{22}=11.38MeV$ and $r_{22\frac{3}{2}}=r_{22\frac{5}{2}}=3.57fm$, $\varepsilon^{s\cdot L}_{nlj}(SSM)$ formula is also very effective.

Table (1 A) shows the evolution process of $O^{15}$ from gas state to liquid state, i. e. from SM to SSM in nuclear shell structure. From $V_C$=-20MeV to $V_C$=-31.7MeV, the $O^{15}$ nucleus is in gas state and its evolution follows NEE(SM). At border nucleus ($V_C$=-2$a_v$=-31.7MeV) the shell structure single particle motion transforms suddenly from SM positive harmonic oscillation to SSM negative harmonic oscillation. Border nucleus energy is $\hbar\omega^{SM}_b=7.5MeV$ and radius is $R_b=4.5fm$. $O^{15}$ border nucleus radius is much larger than $O^{15}$ liquid drop radius $R_{DM}=2.96fm$. $O^{15}$ border nucleus is the real halo nucleus. When nuclear single particle motion changes at border nucleus state, nucleus radius increases and the $\varepsilon^{s\cdot L}_{11\frac{1}{2}}(SSM)$ of single particle orbit $P\frac{1}{2}$ also transforms suddenly from negative value to positive value. From $V_C$=-31.7MeV to $V_C$=-48.3MeV, $O^{15}$ nucleus radius shrinks following SSM evolution liquefaction process. At $V_C$=-48.3MeV, $O^{15}$ nucleus radius equals liquid drop radius and nucleus are in full liquid state and normal $O^{15}$ nucleus is formed. $O^{15}$ nucleus liquefaction process is from border nucleus to normal nucleus. The temperature increases from 0 to 16.63MeV. The root mean square radius for outer shell shrinks from 4.47fm to 2.96fm. In the calculation, $E_{DM}(A,N)$ values in NEE (SM,SSM) are obtained from experiment $O^{15}$ binding energy. Table (1B) shows the evolution process of $L_i^{11}$. SSM negative harmonic oscillation in nuclear evolution process forms the neutron halo of $L_i^{11}$ at ground state. Large asymmetry energy of $L_i^{11}$ stops full liquefaction process in $L_i^{11}$ and neutron halo is formed as ground state. NEE(SM,SSM) gives the formation of neutron halo for $L_i^{11}$ nucleus.

Fig(1) shows $O^{15}$ single particle harmonic oscillation energy levels $\varepsilon_{nl}^{SM}(0)$ in SM evolution process. decreasing of $V_C$ results the increasing of $\hbar\omega^{SM}$ and broadening of oscillating energy level difference. Fig(2) shows $O^{15}$ single particle negative harmonic oscillation of in SSM liquefaction process. The energy level difference $\varepsilon_{nl}^{SSM}(0)$ is equal to $\hbar\omega_b^{SM}$ As $V_C$ decreases in SSM liquefaction process, the energy level difference of $\varepsilon_{nl}^{SSM}(0)$ does not change.

IN SSM evolution from border nucleus to normal nucleus, the summation of coulomb energy, surface energy and asymmetry energy is conserved. For light nucleus such as $O^{15}$ and $N^{15}$, the energy level difference of $\hbar\omega_b^{SM}$ remain about the same. This is due to the fact that energy correction terms $\varepsilon_{nlj}^{s \cdot L}(SSM)$ and $\varepsilon_{nl}^{L}(SSM)$ are small. The experiment measurements of $\hbar\omega^{ex}(N^{15},O^{15}) \approx \hbar\omega_b^{SM}(N^{15},O^{15})$ is a conformation for NEE(SM,SSM). $\hbar\omega^{ex}(N^{15},O^{15}) << \hbar\omega_{Rdm}^{SM}(N^{15},O^{15})$ points to the problem of conventional SM. The physics picture of conventional

SM is correct in gas SM evolution process but is incorrect in SSM liquefaction process. The total evolution process to form normal nucleus at ground state is described by NEE(SM, SSM). For normal nucleus at ground state the density of single particle level in SSM is larger than the density of single particle level in SM and the energy of outside single particle in SM is larger than the energy of outside single particle in SSM. The $V_C$ decrease in SSM is larger than the $V_C$ decrease in SM so that for nuclear system SSM system is more stable than SM system according to the principle of potential energy minimum. The temperature concept in SSM is interesting because it illustrates that nuclear motion includes both deterministic part and stochastic part. SSM fully incorporate DM picture and the nucleus generated by SSM is more stable than nucleus in SM form energy minimization point of view.

For heavy nucleus, because nuclear force is independent on charge, the $V_C$ of NEE (SM , SMM) should be the same for for neutron and proton. However, in SM evolution process, proton $\hbar\omega^{SM}(P)$ is larger than neutron $\hbar\omega^{SM}(N)$ due to the large coulomb energy in heavy nucleus. For the same single particle configuration, $\hbar\omega^{SM}(P) > \hbar\omega^{SM}(N)$ so that the mean square root radius $r_{nl}^{SM}(P)$ of proton is smaller than $r_{nl}^{SM}(N)$ of neutron $r_{nl}^{SM}(P) < r_{nl}^{SM}(N)$ in SM evolution process. Only after crossing border nucleus, proton $\hbar\omega^{SSM}(P)$ is smaller than neutron $\hbar\omega^{SSM}(N)$: $\hbar\omega^{SSM}(P) < \hbar\omega^{SSM}(N)$ and $r_{nl}^{SSM}(P) > r_{nl}^{SSM}(N)$ for the same single particle configuration. This is a very important result. For heavy nucleus the neutron number N is much larger than the proton number Z, N>>Z, and the configuration state of neutron is higher than the configuration states of proton. According to SM, the neutron distribution radius would be much bigger than proton distribution radius. This is inconsistent with liquid drop model. Only after SSM evolution process, proton distribution radius $R_{DM}$(P) equals neutron distribution radius $R_{DM}$(N). This reproduces the physics picture of liquid drop model: $R_{DM}$(P) = $R_{DM}$(N) = $R_{DM}$.

In DM the changing of binding energy $E_{DM}(A,N)$ with (A,N) is smooth but the changing of $\varepsilon_{nlj}^{s \cdot L}(SSM)$ with (A,N) is fluctuating. $E_{DM}(A,N)$ can be corrected by $\sum \varepsilon_{nlj}^{s \cdot L}(SSM)$:

$$E_{DM}^{C}(A,N) = E_{DM}^{C}(A,N) - \sum \varepsilon_{nlj}^{s \cdot L}(SSM) \qquad (34)$$

The calculated results of $\sum \varepsilon_{nlj}^{s \cdot L}(SSM)$ show that the correction of $\sum \varepsilon_{nlj}^{s \cdot L}(SSM)$ in $E_{DM}(A,N)$ corresponds to the shell correction of $E_{DM}(A,N)$. In the calculation of $\sum \varepsilon_{nlj}^{s \cdot L}(SSM)$, because $\varepsilon_{nlj}^{s \cdot L}(j = L \pm 1/2)$ of inside particle orbits cancel each other, only outside particle orbits $\varepsilon_{nlj}^{s \cdot L}$ needs to be considered. According to DM, for normal nucleus, $R_{DM}$(P) = $R_{DM}$(N) = $R_{DM}$, the mean square root radius of outside single particle orbits (neutron and proton) are equal to $R_{DM}$, $r_{nl}^{SSM}(n) = r_{nl}^{SSM}(p) = R_{DM}$. $\omega^{SSM}(n, p)$ of outside single particle orbits can be derived and $\sum \varepsilon_{nlj}^{s \cdot L}(SSM)$ can be calculated without introducing any additional parameter. Table (2) is calculated $\sum \varepsilon_{nlj}^{s \cdot L}(SSM)$ values nuclei for ten nuclei. In order to investigate the effect of Most of selected nuclei are magic nuclei on $E_{DM}(A,N)$, most of nuclei are selected from magic nuclei. The calculation results show that $\sum \varepsilon_{nlj}^{s \cdot L}(SSM)$ correction is the shell correction. $\sum \varepsilon_{nlj}^{s \cdot L}(SSM)$ formula is indeed effective.

Based upon $\hbar \omega_b^{SM}$, $\varepsilon_{nlj}^{s \cdot L}(SSM)$ and $\varepsilon_{nl}^{l}(SSM)$ calculation, the distribution of single particle level $\varepsilon_{nlj}^{SSM}(N,P)$ for neutron and proton can be obtained: $\varepsilon_{nlj}^{SSM} + T = \varepsilon_b^{SM} + \varepsilon_{nlj}^{s \cdot L}(SSM) + \varepsilon_{nl}^{l}(SSM)$. Fig (3) is the single distribution of single particle level $\varepsilon_{nlj}^{SSM}(N,P)$ in SSM. In $\varepsilon_{nlj}^{SSM}(N)$ of Fig(3), there is no single particle level between two levels of j=l $\pm \frac{1}{2}$ except n=2 shell. In $\varepsilon_{nlj}^{SSM}(P)$ of Fig(3), there is also not single particle level between two levels of j=l $\pm \frac{1}{2}$ except n=2 and n=3 shells.

Fig.(4) shows the dependence of total binding energy ($E_{DM}(A,N)$ - $E_{exp}(A,N)$) upon neutron number N. It is also $\sum \varepsilon_{nlj}^{s \cdot L}(SSM)$ versus N curve. The fluctuating $\sum \varepsilon_{nlj}^{s \cdot L}(SSM)$ versus N curve shows that the $\varepsilon_{nlj}^{s \cdot L}(SSM)$ of j=l $+ \frac{1}{2}$ state increases binding energy to form magic nuclei with N=28、50、82、126. And then the $\varepsilon_{nlj}^{s \cdot L}(SSM)$ of j=l $- \frac{1}{2}$ state decreases binding energy. Fig. (3) is consistent with Fig. (4). The nuclear shell structure depends on nuclear evolution history. The magic nuclei with N,P=2、8 and 20 formed in SM evolution process and The magic nuclei with N,P=28,50,82,126 and 184 formed in SSM liquefaction process .In the conventional SM the nuclear shell structure does no depend on historical process of nucleus formation so the conventional SM is not perfect.

The single particle configurations in SM and SSM are same. The deformation shell model corresponding to SSM and cranking shell model corresponding to SSM can be easily obtained by introducing the nuclear deformation and nuclear collective rotation [4] [6] [7] to SSM. It is a nature step to reconstruct more complete theory of nuclear shell structure based upon SSM and NEE in current manuscript.

Finally let us discuss the relationship between nuclear evolution and the universe expansion[8]. The universe came from Big Bang and the universe was expanding rapidly. About 3 min after Big Bang the temperature and the nucleon density dropped steadily. When the nucleon density in the universe was much smaller than one of normal nucleus, the short range strong nuclear force attracts nucleons together to form nucleon gas groups. These nucleon gas groups were separated from each other due to the universe expansion. The nucleon gas groups went through the evolution process from gas state to liquid state (from SM process to SSM process) and formed different nuclei in the universe. The universe continued expanding and the electromagnetic force combined nuclei and electrons to form atoms and molecules. Then Gravity attracted matter together to form galaxies, stars and planets.

**Acknowledgement:** It is pleasure for us to acknowledge some useful discussions with professor Shuwei Xu.

Table (1 A) evolution process of O$^{15}$

| V$_c$(Mev) | - 20 | - 26 | - 31.7 | - 40 | - 48.3 |
|---|---|---|---|---|---|
| T (Mev) | | | 0.00 | 8.33 | 16.63 |
| $\varepsilon_{11}(0)$(Mev) | -14.33 | -13.60 | -12.95 | -21.27 | -29.58 |
| $\varepsilon_{00}(0)$(Mev) | -16.60 | -18.56 | -20.44 | -28.76 | -37.07 |
| $\varepsilon_{11\frac{1}{2}}^{\frac{5}{2}}$(Mev) | -0.03 | -0.12 | -0.27/0.13 | 0.35 | 0.67 |
| $\varepsilon_{11\frac{3}{2}}^{\frac{5}{2}}$(Mev) | 0.01 | 0.06 | 0.13/-0.06 | -0.17 | -0.34 |
| r$_{11}$(fm) | 6.67 | 4.57 | 3.72/4.47 | 3.49 | 2.96 |
| r$_{oo}$(fm) | 5.23 | 3.54 | 2.14/2.88 | 1.80 | 1.58 |

Vc=-2a$_v$=-31.7MeV          R$_{DM}$(O$^{15}$)=2.96fm     E$_{exp}$(O$^{15}$)=112.0MeV

Table (1 B) evolution process of $L_i^{11}$

| $V_c$(Mev) | -20 | -26 | -31.7 | -34 | -40 |
|---|---|---|---|---|---|
| T (Mev) | | | 0.00 | 2.33 | 8.33 |
| $\varepsilon_{11}(0)$(Mev) | -6.285 | -5.264 | -4.30 | -6.629 | -12.63 |
| $\varepsilon_{00}(0)$(Mev) | -11.771 | -13.558 | -15.247 | -17.577 | -23.576 |
| $\varepsilon_{11\frac{1}{2}}^{s,l}$(Mev) | -0.108 | -0.247 | -0.431/0.0106 | 0.0252 | 0.0915 |
| $\varepsilon_{11\frac{3}{2}}^{s,l}$(Mev) | 0.036 | 0.0823 | 0.144/-0.0035 | -0.00832 | -0.0302 |
| $r_{11}$(fm) | 4.356 | 3.534 | 3.076/7.76 | 6.25 | 4.53 |
| $r_{oo}$(fm) | 3.366 | 2.738 | 2.38/2.47 | 2.303 | 1.988 |

$V_c = -2a_v = -31.7$ MeV         $E_{exp}(L_i^{11}) = 45.54$ MeV

Table (2) calculation values of $\sum \varepsilon_{nlj}^{s.l}$ (SSM) for nucleis

| $_zX^N$ | $_{92}U^{138}$ | $_{82}Pb^{126}$ | $_{72}Hf^{112}$ | $_{58}Ce^{92}$ | $_{54}Xe^{82}$ | $_{58}Ce^{72}$ | $_{40}Zr^{58}$ | $_{38}Sr^{50}$ | $_{26}Fe^{28}$ | $_{20}Ca^{20}$ |
|---|---|---|---|---|---|---|---|---|---|---|
| $\sum \varepsilon_{nlj}^{s.l}$ (SSM) | 0 | -11.71 | -0.805 | 0 | -8.92 | -1.27 | 0 | -6.55 | -8.28 | 0 |

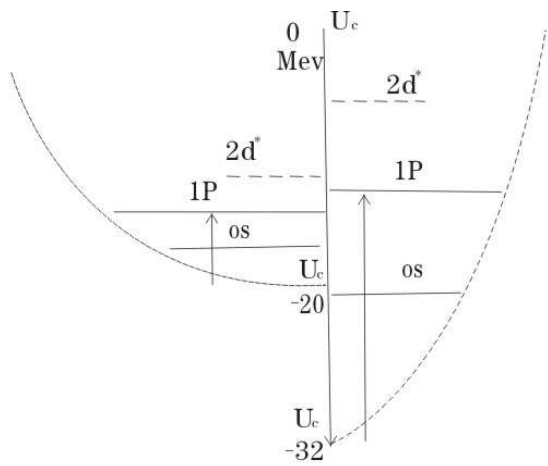

Fig(1) The change of single particle level $\varepsilon_{nl}^{SM}(0)$ with $V_c$ in SM evolution process

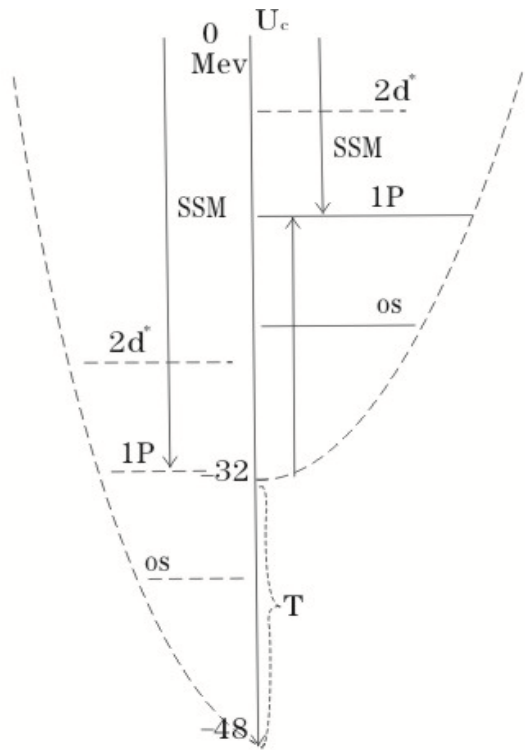

Fig(2) The change of single particle level $\varepsilon^{SSM}_{nl}(0)$ with $V_c$ in SSM Liquefaction process

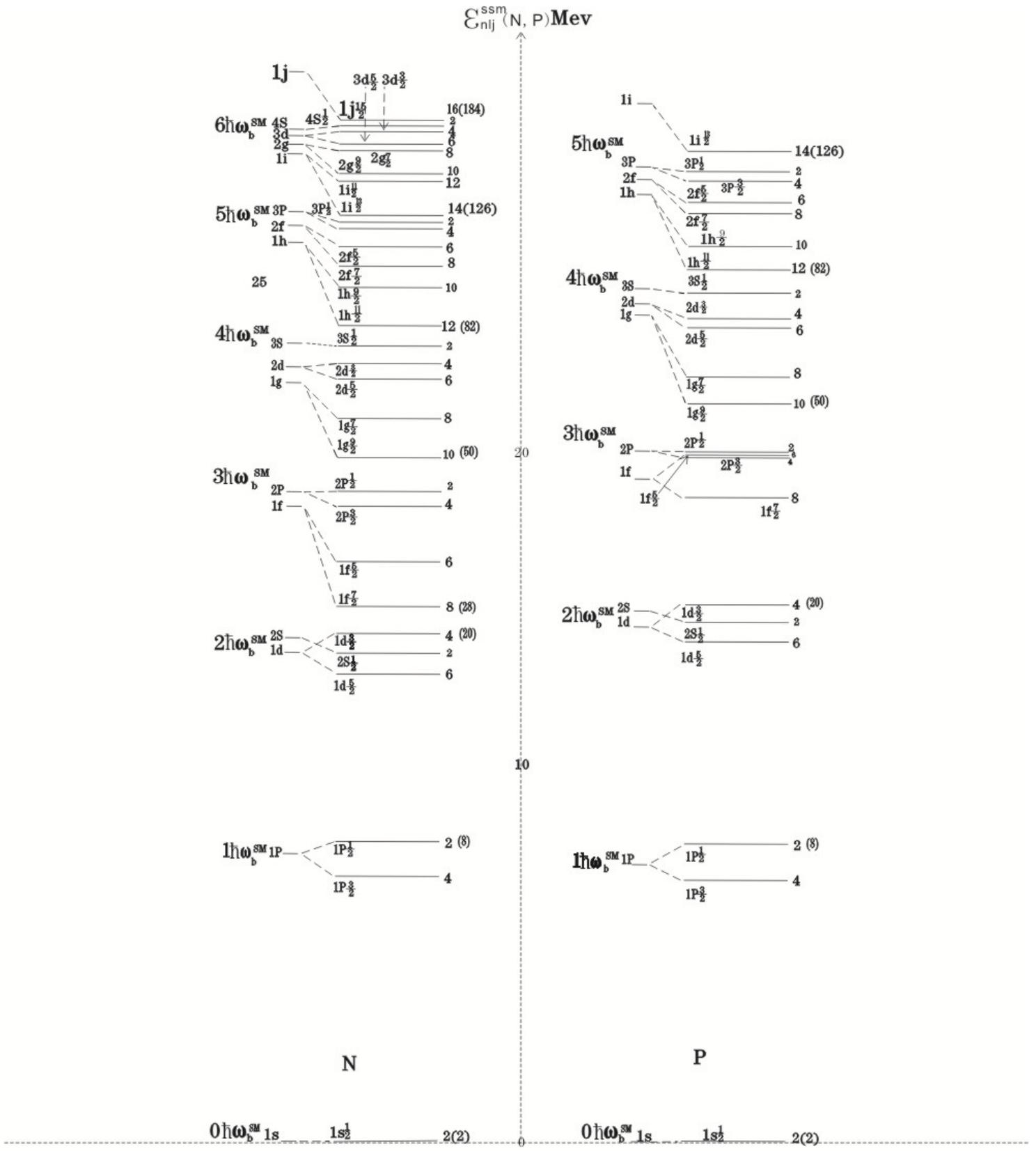

Fig (3) In SSM the distribution of single particle level $\varepsilon_{nlg}^{ssm}(N, P)$

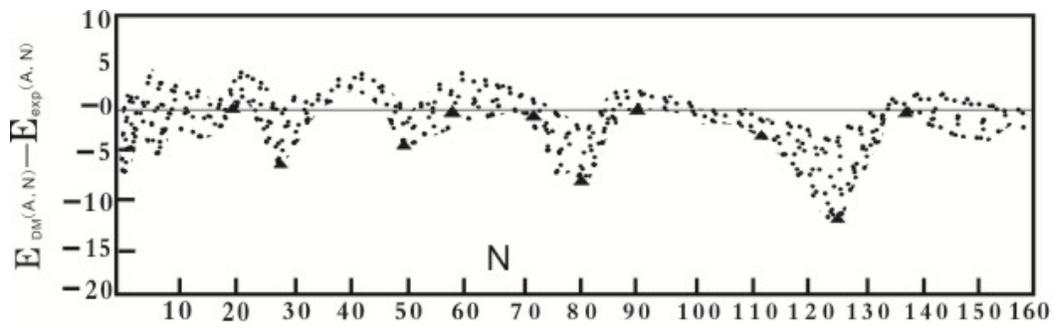

Fig(4) The change curve of nuclear binding energy $E_{DM}(A,N)$ with neutron number N